\documentclass[english,prl,twocolumn,superscriptaddress,amsmath,amssymb,a4paper]{revtex4}
\usepackage[T1]{fontenc}
\usepackage[latin1]{inputenc}
\usepackage{graphicx}
\usepackage[top=0.85in,bottom=0.9in,,left=0.75in,right=0.75in]{geometry}
\usepackage{booktabs}
\usepackage{color}
\usepackage{verbatim}

\makeatletter
\usepackage{dcolumn}
\usepackage{bm}
\usepackage{epsfig}

\newcommand{\ignore}[1]{}

\usepackage{babel}
\makeatother

\usepackage{sidecap}
\sidecaptionvpos{figure}{c}

\begin{document}

\title{Generic Approach to Intrinsic Magnetic Second-order Topological Insulators via Inverted $p-d$ Orbitals}

\author{Zhao Liu}
\email{Zhao.Liu@monash.edu}
\affiliation{Department of Materials Science and Engineering, Monash University, Victoria 3800, Australia}
\affiliation{ARC Centre of Excellent in Future Low-Energy Electronics Technologies, Monash University, Victoria 3800, Australia}

\author{Bing Liu}
\affiliation{School of Physics and Electronic Information, Weifang University, Weifang, Shandong 261061, China}

\author{Yuefeng Yin}
\affiliation{Department of Materials Science and Engineering, Monash University, Victoria 3800, Australia}
\affiliation{ARC Centre of Excellent in Future Low-Energy Electronics Technologies, Monash University, Victoria 3800, Australia}

\author{Nikhil V. Medhekar}
\email{Nikhil.Medhekar@monash.edu}
\affiliation{Department of Materials Science and Engineering, Monash University, Victoria 3800, Australia}
\affiliation{ARC Centre of Excellent in Future Low-Energy Electronics Technologies, Monash University, Victoria 3800, Australia}

\begin{abstract}
The integration of intrinsically magnetic and topologically nontrivial two-dimensional materials holds tantalizing prospects for the exotic quantum anomalous Hall insulators and magnetic second-order topological insulators (SOTIs). Compared with the well-studied nonmagnetic counterparts, the pursuit of intrinsic magnetic SOTIs remains limited. In this work, we address this gap by focusing on $p-d$ orbitals inversion, a fundamental but often overlooked phenomena in the construction of topological materials. We begin by developing a theoretical framework to elucidate $p-d$ orbitals inversion through a combined density-functional theory calculation and Wannier downfolding. Subsequently we showcase the generality of this concept in realizing ferromagnetism SOTIs by identifying two real materials with distinct lattices: 1$T$-VS$_2$ in a hexagonal lattice, and CrAs monolayer in a square lattice. We further compare it with other mechanisms requiring spin-orbit coupling and explore the similarities to topological Kondo insulators. Our findings establish a generic pathway towards intrinsic magnetic SOTIs.  
\end{abstract}

\maketitle
\textit{Introduction.} 
In recent years, the concept of $\mathbb{Z}_2$ topological insulators (TIs) \cite{Kane2005, Bernevig2006} has evolved to encompass higher-order topology \cite{Benalcazar2017a, Benalcazar2017b, Langbehn2017, Song2017, Schindler2018a, Miert2018, Xie2021}, where an $m$-dimensional $n$-th order TI is characterized by gapless states in the $(m-n)$-dimensional subspace. With $n$ = 2, the second order three-dimensional/two-dimensional topological insulator (3D/2D SOTI for $m$ = 3/2) host gapless 1D hinge/0D corner states, respectively. For real applications, 1D hinge states provide a good platform for realizing Luttinger liquid \cite{Wang2023} and 0D corner states for Majorana zero modes \cite{Yan2018, Yan2019}. The hallmark gapless 1D hinge states have been experimentally observed between two gapped surfaces in bulk bismuth \cite{Schindler2018b}, $T_d$-WTe$_2$ \cite{Choi2020, LeeJ2023} and Bi$_4$Br$_4$ \cite{Noguichi2021}. In the realm of 2D SOTI, pioneering studies have focused on "breathing" lattices where intercell and intracell hoppings differ \cite{Motohiko2018}, such as the breathing kagome lattice realized in transition metal dichalcogenide \cite{Jung2022, Zeng2021}, and the breathing Kekule lattice in both graphyne \cite{Sheng2019, Liub2019, Lee2020} and covalent organic frameworks \cite{Hu2022, Ni2022}. The mechanical twisting of trivial 2D materials offers another approach to realize SOTI and has been reported in both twisted bilayer graphene and $h$-BN \cite{Liub2021}. Recently, the unique 0D corner state has been experimentally identified in metal organic framework Ni$_3$(HITP)$_2$ \cite{Hu2023}.

Nevertheless, the requirement of time-reversal symmetry for $\mathbb{Z}_2$ TIs generally excludes the large family of magnetic materials, especially the recently discovered 2D ferromagnetic insulators \cite{Huang2017, Gong2017, Zhang2019, Kong2019, Deng2020, LeeK2021, Li2023, Liu2023}. Since 2D topological materials are attractive for device applications, it is crucial to extend the idea of higher-order topology to 2D magnetic materials. Accordingly, candidate materials such as monolayer $2H$-MX$_2$ (M = Ru, Os and X= Br, Cl) \cite{LuR2023, Cai2023} and CrSiTe$_3$ \cite{WangX2023}, have been proposed as ferromagnetic SOTIs. However, compared to their nonmagnetic counterparts, the underlying mechanisms for realizing 2D magnetic SOTIs remain relatively restricted. The engineering of first-order TI via Zeeman field is expected to provide a universal approach towards magnetic SOTI \cite{Ren2020, Chen2020, Mu2022}, as long as the applied Zeeman field does not alter the band ordering protected by spin-orbit coupling (SOC). Nevertheless, this condition is always violated in intrinsic ferromagnetic insulators where the exchange field ($\sim$ 1.0 eV) determined by electron-electron interactions significantly outweighs SOC ($\sim$ 0.1 eV).

Physically, band inversions due to orbitals with different parities have played a crucial role in achieving first-order topology. In topological Kondo insulators, when the strong electron-electron interactions push up $f$ orbitals into a $d$ band, band inversion takes place, triggering a topological phase transition \cite{Dzero2010, Dzero2016}. In noble metals like Au and Ag, the outermost $s$ orbital has a high energy with respect to $p$ orbitals, giving rise to a topological origin to Shockley surface states \cite{Yan2015, Zhang2017}. Transition metal compounds (TMCs) present another scenario where metal cation's $d$ and the ligand's $p$ orbitals naturally exhibit opposite parities. For example, the band inversion occurs between $p$ and $d$ orbitals have lead to quantum spin Hall insulators in $1T'$ transition metal dichalcogenides \cite{Qian2014}. Considering the fact TMCs often exemplify correlated materials, the correlated charge gap tends to be topologically trivial. However, as we illustrate below, correlated TMCs can harbor a hidden $p-d$ orbitals order termed as \textit{$p-d$ orbitals inversion}. This phenomenon, as fundamental as electron-electron correlations, can engender higher-order topology without the aid of SOC. 

\begin{figure}
\includegraphics[width=8cm]{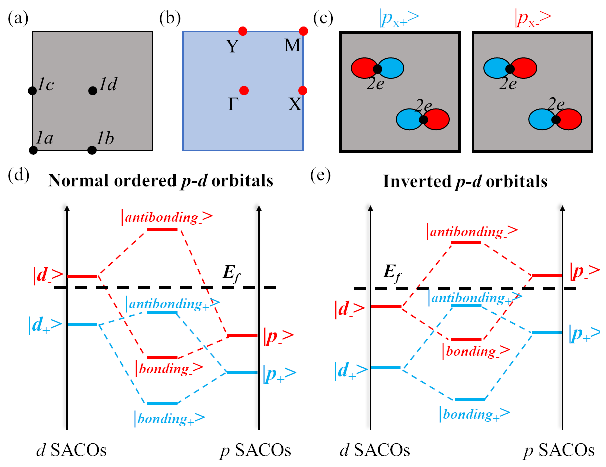}
\caption{(a) Four high symmetric WPs in 2D $\mathcal{P}$-symmetric lattice: the fractional coordinates are $1a$ = (0.0, 0.0), $1b$ = (0.5, 0.0), $1c$ = (0.0, 0.5) and $1d$ = (0.5, 0.5). (b) Four TRIM in the first BZ: $\Gamma$, $X$, $Y$, and $M$. (c) The linear combination of one $p_x$ orbital at WP $2e$ gives SACOs $|p_{x\pm}>$ with parity "$\pm$". Orbital interaction diagram for (d) normal ordered $p-d$ orbitals and (e) inverted $p-d$ orbitals. Blue/red color is used to mark parity $\pm$.}
\label{FIG-1}
\end{figure}

\begin{figure*}
\includegraphics[width=14cm]{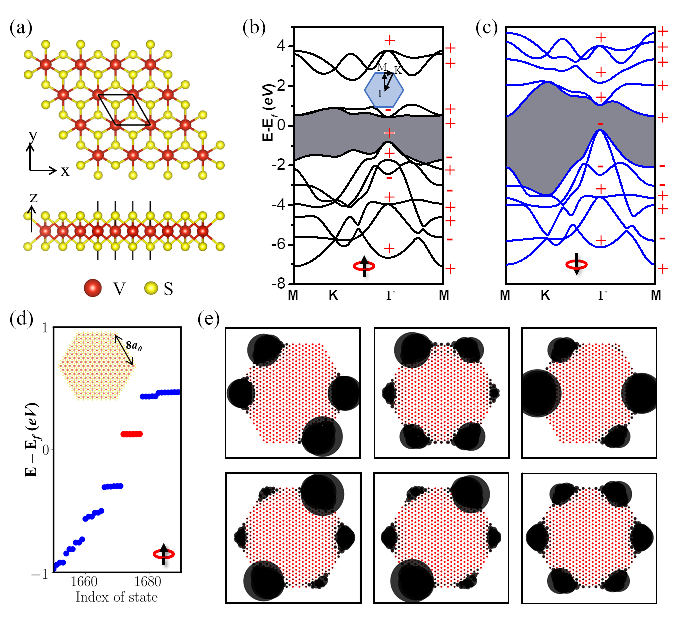}
\caption{(a) Top and side view of 1$T$-VS$_2$ monolayer, the center of inversion locates at V site. Band structure of hybrid functionals for (b) spin up and (c) spin down channel. Label "+/-" indicates (degenerated) Bloch states are parity even/odd. The band gap is shaded by grey color. (d) Energy spectrum of for a hexagonal cluster (inset) in the spin up channel. (e) The spatial charge distributions of the six states marked by red dots in (d).}
\label{FIG-2}
\end{figure*}

\begin{figure*}
\includegraphics[width=14cm]{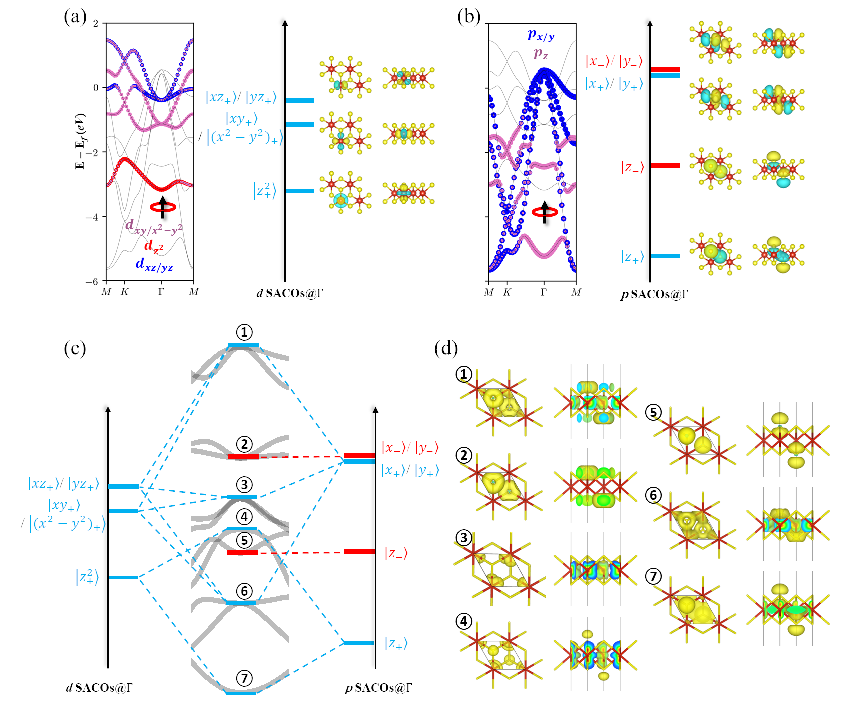}
\caption{ Determination of (a) $d$ SACOs and (b) $p$ SACOs at $\Gamma$ in 1$T$-VS$_2$ monolayer. (c) Orbital interaction diagram, the curve in the middle is the band dispersion near $\Gamma$. (d) Top and side view of the charge densities for Bloch states labelled as \textcircled{1}-\textcircled{7} in (c). For degenerated states, only one charge density is plotted.}
\label{FIG-3}
\end{figure*} 

\textit{Topological invariant based on parity.}  Topological invariant is effective in diagnosing topological phases. As intrinsic magnetic insulators commonly occupy inversion symmetry $\mathcal{P}$, here we consider $\mathcal{P}$-based topological invariant \cite{Miert2018, Po2017, Kruthoff2017}. A $\mathcal{P}$-invariant 2D unit cell is displayed in {\color{blue} Fig. \ref{FIG-1} (a)}, together with the four Wyckoff positions (WPs) $1a$, $1b$, $1c$, and $1d$ whose site symmetry including $\mathcal{P}$. In duality, four time-reversal invariant momentum (TRIM) $\Gamma$, $X$, $Y$ and $M$ exist in the first Brillouin zone (BZ) as shown in {\color{blue} Fig. \ref{FIG-1}(b)}. Denoting $\Gamma_{+/-}, X_{+/-}, Y_{+/-}$, and $M_{+/-}$ as the number of occupied Bloch states with odd/even parity at the four TRIM, the electric dipole ($p_x, p_y$) and quadrupole polarization ($q_{xy}$) are calculated as \cite{SI}:
\begin{equation}
\begin{split}
(p_x, p_y) &= (\frac{\Gamma_- -  X_-}{2}, \frac{\Gamma_- -  Y_-}{2}) \mod 1 \\
q_{xy} &=  \frac{\Gamma_- -  X_- - Y_- + M_-}{8} \mod \frac{1}{2} \\
\end{split}
\label{eq:dipole-quad}
\end{equation}
Since a nonvanishing $q_{xy}$ requires vanishing dipole polarization, following {\color{blue}Eq.\ref{eq:dipole-quad}}, the condition for a SOTI phase is
\begin{equation}
\begin{split}
&(\Gamma_- -  X_-) \mod 2 = (\Gamma_- -  Y_-) \mod 2 = 0 \\  
&\frac{\Gamma_- -  X_- - Y_- + M_-}{8} \mod \frac{1}{2} \neq 0 \\
\end{split}
\label{eq:condition-QI}
\end{equation}
If both $(\Gamma_- -  X_-)$ and $(\Gamma_- -  Y_-)$ are odd rather than even, first-order TIs is preferred via the Fu-Kane formula \cite{Fu2007}. Therefore {\color{blue}Eq.\ref{eq:condition-QI}} puts strong constraints on the numbers of band inversions in realizing SOTI. For spin polarized materials, we can treat each spin channel separately, mainly due to the linearity of the topological invariant. 

\begin{figure*}
\includegraphics[width=17cm]{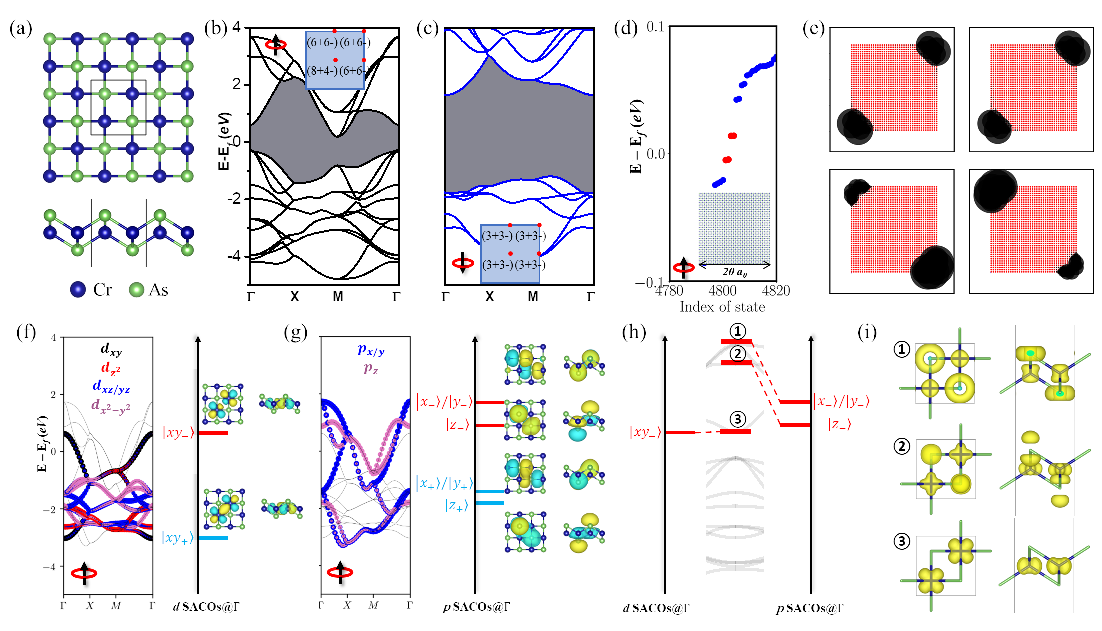}
\caption{ (a) Top and side view of CrAs monolayer, the center of inversion locates at the middle of two Cr sites. Band structure of hybrid functionals for (b) spin up and (c) spin down channel with parity distributions. The label "(8+4-)" at $\Gamma$ means 8/4 valence states have even/odd parity. (d) Energy spectrum of for a square cluster (inset) in spin up channel. (e) The spatial charge distributions of the four states marked by red dot in (d). (f) Determination of $d$ SACOs $|xy_{\pm}>$ at $\Gamma$ point. All the left $d$ SACOs are topological inert and ignored. (g) Determination of $p$ SACOs at $\Gamma$. (h) Orbital interaction diagram, the curve in the middle is the band dispersion around $\Gamma$. For simplicity, only SACOs related to band inversions are plotted. (i) Top and side view of the charge densities for Bloch states labelled by \textcircled{1}-\textcircled{3} in (h). }
\label{FIG-4}
\end{figure*}

\textit{Characterization of $p-d$ orbitals inversion.} 
TMCs are described by the multi-band $d-p$ model:
\begin{equation}
\hat{H} = \hat{H}_d + \hat{H}_p + \hat{H}_{dp}
\label{eq:Hamiltonian}
\end{equation}
where $\hat{H}_d$ ($\hat{H}_p$) describes $d$ ($p$) orbitals of metal cations (ligand anions) and $\hat{H}_{dp}$ contains interactions between $p$ and $d$ orbitals, like the $p-d$ orbitals hybridization. It is assumed that the strong correlations in {\color{blue}Eq.\ref{eq:Hamiltonian}} give an ordered phase that can be captured by a mean-field theory, such as the long-range ferromagnetism discussed herein. Therefore, magnetic Wannier orbitals serve as a natural basis. 

For magnetic Wannier orbitals at the same WPs, we can take their linear combinations to form cluster orbitals with either odd or even parity at TRIM, which we dubbed as symmetry-adapted cluster orbitals (SACOs). {\color{blue} Fig. \ref{FIG-1} (c)} shows the two $p$ SACOs $|p_{x\pm}>$ ($\pm$ means parity even/odd) constructed from the $p_x$ orbital at WP $2e$. Since intracell hopping is always larger than that of intercell, $|p_{x+}>$ will always has lower energy than $|p_{x-}>$. Obviously both $d$ and $p$ orbitals can form SACOs and according to their onsite energy differences, we have at least two cases: (1) if all $p$ SACOs are occupied at all TRIM as shown in {\color{blue} Fig. \ref{FIG-1} (d)}, we have normal ordered $p-d$ orbitals, which is shared by most TMCs; (2) if at certain TRIM, there exists unoccupied $p$ SACO which is higher than all $d$ SACOs as shown in {\color{blue} Fig. \ref{FIG-1} (e)}, it is called inverted $p-d$ orbitals. With further inclusion of $p-d$ orbital hybridization, if $p$ and $d$ SACOs share the same parity, they will repeal each other and form gap sandwiched by bonding and antibonding SACOs $|bonding_{\pm}>/|antibonding_{\pm}>$, as sketched in the orbital interaction diagram {\color{blue} Fig. \ref{FIG-1} (d)-(e)}. The square of $|bonding_{\pm}>/|antibonding_{\pm}>$ gives the charge densities which are accessible through density-functional theory (DFT) calculations. Nevertheless, when examining charge densities alone, the difference between the orbital orders represented by {\color{blue} Fig. \ref{FIG-1} (d)} and {\color{blue} Fig. \ref{FIG-1} (e)} is always ignored, as both are contributed by $p$ and $d$ orbitals. To recover the information about the orbital order, it is necessary to switch off $p-d$ orbital hybridization. Since DFT can not switch off $\hat{H}_{dp}$, here we take advantage of Wannier downfolding: Wannier downfolding is firstly conducted using all $d$ and $p$ orbitals to derive an effective Hamiltonian $\hat{H}_{eff}$ approximating $\hat{H}$. Then, by setting all $p-d$ orbital hybridizations to zero, $\hat{H}_{eff}$ is reduced to a diagonal form $\hat{H}_{d,eff} + \hat{H}_{p,eff}$, allowing for the extraction of onsite energies of $p$ and $d$ SACOs at all TRIM.  In this way, the $p-d$ orbital order can be mapped out. It is emphasized that $p-d$ orbitals inversion can be regarded as a real space manifestation of band inversion defined in the reciprocal momentum space. Notably, the $\mathcal{P}$ symmetry applied here can be replaced by any other crystalline symmetry. Therefore, $p-d$ orbitals inversion does not inherently depend on specific crystalline symmetries and should manifest as a generic phenomenon across various crystalline systems. In the following, we will illustrate how inverted $p-d$ orbitals give rise to intrinsic ferromagnetic SOTIs in the 1$T$-VS$_2$ monolayer in a hexagon lattice and the CrAs monolayer in a square lattice. 

\textit{Hexagonal lattice.} 
The 2D 1$T$-VS$_2$ shares a hexagon lattice as depicted in {\color{blue} Fig. \ref{FIG-2} (a)}. Being a prototypical $d^1$ system, 1$T$-VS$_2$ is anticipated to display phenomena such as charge density wave (CDW) and ferromagnetism. However, there are controversies regarding CDW and ferromagnetism in both experiments and theoretical studies. In bulk 1$T$-VS$_2$, magnetic \cite{Murphy-1977-exp} and nuclear magnetic resonance \cite{Tsuda-1983-exp} measurements have indicated a CDW transition below 305 K, while high-pressure samples have not detected CDW down to 94 K \cite{Gauzzi-2014-exp}. For monolayers, Arnold et al. \cite{Arnold-2018-exp} found no evidence of CDW on Au(111) surface; however Efferen et al. \cite{Efferen-2021-exp} observed a CDW gap above the Fermi level on graphene/Ir(111) surface. Predicted to be a ferromagnetic metal \cite{Ma-2012-theo} without CDW \cite{Issacs-2016-theo, Zhuang-2016-theo}, Gao et al. \cite{Gao-2013-exp} reported room temperature ferromagnetism in ultrathin 1$T$-VS$_2$ nanosheets by observing magnetic hysteresis and ferromagnetic resonance. Subsequently, Guo et al. \cite{Guo-2017-exp} revealed that 1$T$-VS$_2$ undergoes a ferromagnetic semiconductor to ferromagnetic metal transition with respect to van der Waals gaps. Here we concentrate on the ferromagnetic semiconducting phase. 

Employing DFT calculations with hybrid functionals \cite{SI}, {\color{blue} Fig. \ref{FIG-2} (b)-(c)} suggest 1$T$-VS$_2$ is a small gap ferromagnetic semiconductor, in consistent with the experiment \cite{Guo-2017-exp}. Although the long-range ferromagnetic order breaks all vertical mirror symmetries, the rotoinversion $S_{6z}$ is conserved and $X$, $Y$, and $M$ share the same parity distribution, thus {\color{blue}Eq.\ref{eq:condition-QI}} now is 
\begin{equation}
\Gamma_- - M_- = 2 \mod 4
\label{eq:condition-QI-S6z}
\end{equation}
According to the parity eigenvalues listed in {\color{blue} Fig. \ref{FIG-2} (b)-(c)}, $\Gamma_- = 1, M_- = 3$ ($\Gamma_- = M_- = 3$) in the spin up (down) channel gives a SOTI (trivial) phase. Due to the bulk-boundary correspondence in SOTIs, the nonvanishing $q_{xy}$ in the spin up channel implies the existence of spin-polarized gapped edge states (see  {\color{blue}Fig. S2} \cite{SI}) and corner states. Given that TMD nanoflakes always adopt hexagonal and triangular shapes during growth, {\color{blue} Fig. \ref{FIG-2} (d)} shows the energy spectrum of a hexagon cluster with an edge length of 8$a_0$ ($a_0$ = 3.23 \AA , the optimized in-plane lattice constant of 2D 1$T$-VS$_2$, in good agreement with experiment \cite{Gauzzi-2014-exp}). It is clear that there are six isolated states whose charge densities are localized at the six corners, reflecting the nontrivial topology in the bulk. Such topological corner states can also be found in triangular shapes \cite{SI}.  

The topological distinction between spin up and down channel can be readily understood through the underlying $p-d$ orbitals order. As shown in {\color{blue}Fig. S1} \cite{SI}, spin down channel has a normal ordered $p-d$ orbitals, where the six valence (five conduction) bands in {\color{blue} Fig. \ref{FIG-2} (c)} originate from the $p$ orbitals of S ($d$ orbitals of V). In other words, the spin down channel is adiabatically connected to an atomic insulator (AI) formed by the S sublattice. This explains why the spin down channel exhibits a parity distribution $\Gamma_- = \Gamma_+ = M_- = M_+ = 3$ in {\color{blue} Fig. \ref{FIG-2} (c)}, and must be topological trivial. Conversely, the situation in the spin up channel varies significantly, particularly at $\Gamma$ when comparing parity distribution in {\color{blue} Fig. \ref{FIG-2} (b)} and {\color{blue} Fig. \ref{FIG-2} (c)}. As illustrated in {\color{blue} Fig. \ref{FIG-3} (a)-(b)}, there is a $p-d$ orbitals inversion at $\Gamma$: the $p$ SACOs $|x_{\pm}>/|y_{\pm}>$ are higher than all the $d$ SACOs. When $p-d$ orbital hybridization is taken into account, $|x_{-}>/|y_{-}>$ becomes (degenerate) nonbonding states \textcircled{2} in {\color{blue} Fig. \ref{FIG-3} (c)}, where the charge densities are entirely contributed by $p_{x/y}$ orbitals, as shown in {\color{blue} Fig. \ref{FIG-3} (d)}. Since such (degenerate) nonbonding states are the lowest conduction states at $\Gamma$, the spin up channel now only has $\Gamma_- = 1$, in contrast to $\Gamma_- = 3$ in the spin down channel. This discrepancy leads to nontrivial topology in the spin up channel as $M_- = 3$ for both spin channels. 

\textit{Square lattice.} 
CrAs monolayer, featuring an anti-PbO type structure shown in {\color{blue} Fig. \ref{FIG-4}(a)}, is predicted to be a high Curie temperature ferromagnetic semiconductor with a good stability \cite{Liu2023, Ma2020}. The pronounced ferromagnetism in this system is demonstrated to be mediated by the significant number of $p$ holes on As atoms \cite{Liu2023}. As we elucidate later, the emergence of such non-negligible $p$ holes is a direct consequence of $p-d$ orbitals inversion.

As CrAs monolayer shares a nonsymmorphic space group $p\frac{4}{n}mm$, the Bloch states at $X$, $Y$, $M$ are two-fold degenerate and each degeneracy shares opposite parity eigenvalues \cite{SI}. Under such circumstances, {\color{blue}Eq.\ref{eq:condition-QI}} becomes:
\begin{equation}
\Gamma_- -  \Gamma_+ = 4 \mod 8
\label{eq:condition-QI-CrAs}
\end{equation}
which implies the construction of SACOs can be narrowed down to $\Gamma$.   

{\color{blue} Fig. \ref{FIG-4}(b)-(c)} presents the DFT band structures based on hybrid functionals and the parity distributions for the 12 (6) valance bands in the spin up (down) channel. The parity distribution $\Gamma_{-} = \Gamma_{+} = 3$ in the spin down channel indicates a trivial phase according to  {\color{blue}Eq.\ref{eq:condition-QI-CrAs}}. In contrast, for the spin up channel, we observe $\Gamma_{-} - \Gamma_{+} = -4$, suggesting a SOTI phase. {\color{blue} Fig. \ref{FIG-4}(d)} displays the energy spectrum of a square cluster with an edge length of $20 a_0$ ($a_0$ = 4.20 \AA). Notably, there are four isolated states with charge localized at the four corners, signifying a nonvanishing $q_{xy}$ in the spin up channel. 

Next we turn to the orbital order. In the spin down channel, the $p-d$ orbitals exhibit a normal ordering \cite{SI} which is continuously connected to the AI formed by the As sublattice, akin to the scenario observed in 1$T$-VS$_2$ monolayer. In contrast, the $p-d$ orbitals are inverted in the spin up channel, as illustrated in {\color{blue} Fig. \ref{FIG-4}(f)-(g)}. Here all antibonding $p$ SACOs $|x_{-}/y_{-}/z_{-}>$ are higher than all $d$ SACOs, contributing to $\Gamma_{-} - \Gamma_{+} = -3$. Additionally, the $d$ SACO $|xy_{-}>$ also plays a crucial role as it constitutes the conduction state \textcircled{3} at $\Gamma$, as illustrated in {\color{blue} Fig. \ref{FIG-4}(h)-(i)}. With all the left $d$ SACOs occupied, $d$ SACO $|xy_{-}>$ contributes $\Gamma_{-} - \Gamma_{+}= -1$. Combining $p$ and $d$ SACOs together, we obtain $\Gamma_{-} - \Gamma_{+} = -4$  in the spin up channel, in agreement with the parity distribution observed in {\color{blue} Fig. \ref{FIG-4}(b)}. 

\textit{Discussions.} 
In the 1$T$-VS$_2$ monolayer, the $d$ SACOs are topologically inert as the parity eigenvalues at $\Gamma$ and $M$ are always the same. Consequently, band inversion can only occur through the empty $p$ SACOs $|x_{-}>/|y_{-}>$. In the CrAs monolayer, the $d$ SACOs do participate in band inversion like $|xy_{-}>$, it seems that SOTI can be realized without the aid of $p-d$ orbitals inversion. However this assumption is incorrect. In an insulator without $p-d$ orbitals inversion, the $p$ orbitals form a closed shell with all $p$ SACOs occupied. To achieve a nonvanishing $q_{xy}$, according to {\color{blue}Eq.\ref{eq:condition-QI-CrAs}}, we must select another six occupied states from the ten $d$ SACOs to satisfy $\Gamma_{-} - \Gamma_{+} = 4 \mod 8$. Nevertheless this is impossible as only $|xy_{-}>$ and $|z^2_{-}>$ possess higher energy than $|xy_{+}>$ and $|z^2_{+}>$ among all the $d$ SACOs. Therefore the maximum value $\Gamma_{-} - \Gamma_{+}$ can reach is 2, as found in the configuration $(xy_-)^1(xy_+)^1(z^2_+)^1(xz_-)^1(yz_-)^1((x^2-y^2)_-)^1$. Hence, the empty $p$ SACOs are necessity for achieving SOTI in the spin up channel, underscoring the significance of inverted $p-d$ orbitals in the CrAs monolayer. 

In $\mathbb{Z}_2$ TIs, SOC is necessary to open a nontrivial gap. In magnetic SOTIs via $p-d$ orbitals inversion, the gap is created through $p-d$ orbital hybridization. Therefore, the magnetic SOTIs studied here are robust to SOC as long as SOC does not alter the orbital order. Given that the gap at $\Gamma$ in the spin up channel is significantly larger than the strength of SOC in both 1$T$-VS$_2$ and CrAs monolayers, the SOTI phase should persist even when SOC is considered, as confirmed by our calculations \cite{SI}. Thus, magnetic SOTIs originating from $p-d$ orbitals inversion represent another class of SOC-free topological phases \cite{Alexandradinata2014}, distinct from the universal approach \cite{Ren2020, Chen2020} where SOC is indispensable. We would like to highlight the similarity between magnetic SOTIs generated by $p-d$ orbitals inversion and topological Kondo insulators. In both topological phases, band inversion is induced by orbital inversions under electron-electron interactions and gap is created via orbital hybridization.

Where to find materials with $p-d$ orbitals inversion ? It is suggested that TMCs with high formally valance states, such as V$^{4+}$ and Cr$^{3+}$ studied in this work, should demonstrate $p-d$ orbitals inversion \cite{Khomskii2014}. With inverted $p-d$ orbitals, electrons tend to transfer from $p$ to $d$ orbitals to avoid high formally valance states, resulting in a surplus of holes in the ligands. These holes is capable of stabilizing strong ferromagnetic order \cite{Khomskii1997, Liu2023}, thus illustrating the dual roles of $p-d$ orbitals inversion in achieving both ferromagnetism and nontrivial topology. This is how strong correlation and nontrivial topology reconcile in this type of materials. It is worth noting that a broad range of TMDs possess high formally valance states and should represent potential candidates for realizing $p-d$ orbitals inversion, as suggested by a recent study \cite{Cheng2024}.

In conclusion, this study unveils that $p-d$ orbitals inversion provides a generic approach to SOTIs in intrinsic ferromagnetic insulators, and demonstrates that ferromagnetic 1$T$-VS$_2$ and CrAs monolayer exhibiting inverted $p-d$ orbitals can indeed host a robust SOTI phase. These findings not only advance the current understanding of magnetic SOTIs but also open up a novel avenue for exploring nontrivial topological phases in correlated materials, like SOTI phase in Kondo lattice \cite{Dzero2016}.

Z. L., Y. Y. and N. V. M. gratefully acknowledge the support from National Computing Infrastructure, Pawsey Supercomputing Facility and the Australian Research Council's Centre of Excellence in Future Low-Energy Electronic Technologies (CE170100039). B. L. is funded by the National Natural Science Foundation of China under Grant No. 12204356 and the Natural Science Foundation of Shandong Province under Grant No. ZR2022QA024.

\end{document}